\documentclass[12pt]{article}
\usepackage{epsfig,alltt}

\usepackage{multirow}
\usepackage{axodraw}

\newcommand{\ep}{\varepsilon}

\newcommand{\ovl}{\overline}

\newcommand{\beq}{\begin{equation}}
\newcommand{\eeq}{\end{equation}}
\newcommand{\bea}{\begin{eqnarray}}
\newcommand{\eea}{\end{eqnarray}}

\newcommand{\ice}[1]{\relax}

\begin{document}

\begin{titlepage}
\rightline{\footnotesize TTP-Number: TTP10-19}
\rightline{\footnotesize SFB-Number: SFB/CPP-10-25}

\vspace{2cm}

\bigskip
\begin{center}
{{\large\bf Four Loop Massless Propagators: a Numerical   Evaluation
	 of All Master Integrals} \\
\vglue 5pt \vglue 1.0cm
{\large  A.V. Smirnov}\footnote{E-mail: asmirnov80@gmail.com}\\
\baselineskip=14pt
\vspace{2mm}
{\normalsize Scientific Research Computing Center, Moscow State
University, 119992 Moscow, Russia
}\\
\baselineskip=14pt
{\large   M. Tentyukov}\footnote{E-mail:
  tentukov@particle.uni-karlsruhe.de}\\
\baselineskip=14pt
\vspace{2mm}
{\normalsize
Institut f\"ur Theoretische Teilchenphysik, Karlsruhe
  Institute of Technology (KIT), D-76128 Karlsruhe, Germany
}
\vspace{2mm} \vglue 0.8cm {Abstract}}
\end{center}
\vglue 0.3cm {\rightskip=3pc
 \leftskip=3pc
\noindent We present numerical results which are needed to evaluate all non-trivial 
 master integrals for 
four-loop massless propagators, confirming 
the recent analytic results of 
\cite{BaCh}
and evaluating an extra
order in $\ep$ expansion for each master integral.
 \vglue 0.8cm}
\end{titlepage}

\section{Introduction}

Sector decomposition in its practical aspect is a constructive method
used to evaluate Feynman integrals numerically. The goal of sector
decomposition is to decompose the initial integration domain into
appropriate subdomains (sectors) and introduce, in each sector, new
variables in such a way that the integrand factorizes, i.e. becomes
equal to a monomial in new variables times a non-singular function.

Originally it was used as a tool for analyzing the convergence and
proving theorems on renormalization and asymptotic expansions of Feynman
integrals \cite{Hepp,theory,BM,BdCMPo,books1a}. 
%
After a pioneering work \cite{BH} 
sector decomposition has become an efficient tool for numerical evaluating Feynman
integrals (see Ref.~\cite{Heinrich} for a recent review).
At present, there are two public codes performing the
sector decomposition \cite{BognerWeinzierl} and \cite{FIESTA}. 


The latter one was named {\tt FIESTA} which stands for ``Feynman Integral
Evaluation by a Sector decomposiTion Approach''. Last year
{\tt FIESTA} has been greatly improved in various aspects
\cite{FIESTA2}. 
The code is
capable of evaluating many classes of integrals that one would not be
able to evaluate with the original {\tt FIESTA 1}. Moreover the code
can now be applied to solve the problem of obtaining asymptotic
expansions of Feynman integrals in various limits of momenta and
masses and to find a list of all poles of an integral in space-time
dimension $d$. During the last year {\tt FIESTA} was widely used,
some of application are listed in \cite{FIESTA-appl}.

In the current paper we present numerical results for {\it master
  integrals} (MI's) for four-loop massless propagators which are
relevant for many important physical applications, like the
calculation of the total cross-section of $e^+ e^- $ annihilation into
hadrons, the Higgs decay rate into hadrons, the semihadronic decay
rate of the $\tau$ lepton and the running of the fine structure
coupling constant (see \cite{Chetyrkin:1996ia,Steinhauser:2002rq} for
details).  We  confirm numerically  the recent analytic results of 
work \cite{BaCh} and evaluate an extra order in epsilon expansion for each MI.

\section{Theoretical background and software structure}

{\tt FIESTA} calculates Feynman integrals with the sector decomposition approach.
It is based on the $\alpha$-representation of Feynman integrals.
After performing Dirac and Lorentz algebra one is left with a scalar dimensionally regularized Feynman integral \cite{dimreg}
\begin{eqnarray}
  F(a_1,\ldots,a_n) &=&
  \int \cdots \int \frac{\mbox{d}^d k_1\ldots \mbox{d}^d k_l}
  {E_1^{a_1}\ldots E_n^{a_n}}\,,
  \label{FI}
\end{eqnarray}
where $d=4-2\ep$ is the space-time dimension, $a_n$ are
indices, $l$ is the number of loops and $1/E_n$ are propagators. We
work in Minkowski space where
the standard propagators are the form $1/(m^2 -p^2-i0)$.
Other propagators are permitted, for example,
$1/(v\cdot k\pm i0)$ may appear in diagrams
contributing to static quark potentials or in
HQET\footnote{{\bf H}eavy-{\bf Q}uark {\bf E}ffective {\bf T}heory} 
where $v$ is the quark velocity
(see, e.g  \cite{Grozin:2004yc}).
Substituting
\begin{eqnarray}
    \frac{1}{E_i^{a_i}}=\frac{e^{ai\pi/2}}{\Gamma(a)}\int_0^\infty \mbox{d}\alpha \alpha^{a_i-1} e^{-iE_i\alpha},
\label{AlphaSubstitution}
\end{eqnarray}
changing the integration order, performing the integration over loop momenta,
replacing $\alpha_i$ with $x_i \eta$ and integrating over $\eta$
one arrives at the following formula (see e.g. \cite{Smirnov}):
\begin{eqnarray}
\nonumber
    &&F(a_1,\ldots,a_n) =
   \\
    &&\frac{\Gamma(A-l d/2)}{\prod_{j=1}^n \Gamma(a_j)}
            \int_{x_j\geq 0} d x_i\ldots d x_{n} \delta\left(1-\sum_{i=1}^n x_i \right)
                \left(\prod_{j=1}^n x_j^{a_j-1}\right) \frac{U^{A-(l+1)d/2}}{F^{A-ld/2}},
\label{Alpha}
\end{eqnarray}
where $A=\sum_{i=1}^n a_n$ and
$U$ and $F$ are constructively defined polynomials of $x_i$.
The formula (\ref{Alpha}) has no sense if some of the indices are non-positive integers,
so in case of those the integration is performed according to the rule
\[
\int_{0}^{\infty}d x \frac{x^{(a-1)}}{\Gamma(a)} f(x) = f^{(n)} (0)
\]
where $a$ is a  non-positive integer.

After performing the decomposition of the integration region into the so-called
\textit{primary sectors} \cite{BH} and making a variable replacement, one results
in a linear combination of integrals of the following form:
\begin{eqnarray}
 \int_{x_j=0}^1 d x_i\ldots d x_{n'}\left(\prod_{j=1}^{n'} x_j^{a_j-1}\right) \frac{U^{A-(l+1)d/2}}{F^{A-ld/2}}
\label{Cube}
\end{eqnarray}

If the functions $\frac{U^{A-(l+1)d/2}}{F^{A-ld/2}}$ had no singularities in $\ep$,
one would be able to perform the expansion in $\ep$ and perform the numerical integration
afterwards. However, in general one has to resolve
the singularities first, which  is not possible for general $U$ and
$F$. Thus, one
starts a process the sector decomposition aiming to end with a sum
of similar expressions, but with new functions $U$ and $F$ which have no singularities (all the singularities
are now due to the part $\prod_{j=1}^n {x'}_j^{a'_j-1}$). Obviously it is a good
idea to make the sector decomposition process constructive and to end
with a minimally possible number of sectors. The way sector decomposition is performed
is called a \textit{sector decomposition strategy} and is an essential
part of the algorithm.

After performing the sector decomposition one can resolve the singularities by
evaluating the first terms of the Taylor series:
in those terms one integration is taken analytically, and the remainder
has no singularities. Afterwards the $\ep$-expansion can be performed
and finally one can do the numerical integration and return the result.

Please keep in mind
that this approach works only using numerical integration:
numeric  values for all invariants should be specified  at the very early stage,
after generating the functions $U$ and $F$.

{\tt FIESTA} is written in {\tt Mathematica} \cite{math7}  and C.  The user
is not supposed to use the C part directly as it is launched from {\tt
  Mathematica} via the Mathlink protocol. When the integrand expressions are ready,
Mathematica submits long strings representing integrands for integration; 
the C part translates them into an internal representation optimizing
evaluation speed. Afterwards it uses some numerical integrator to 
perform the numerical integration of the integrand. 
The original {\tt FIESTA} employed  a {\tt Fortran} implementation of {\tt Vegas} 
as an integrator. Later we plugged in the {\tt Cuba} library \cite{Cuba}. By
default {\tt FIESTA} uses the {\tt Vegas} integrator, but this
behavior can be easily controlled by the user. Both Mathematica and C
parts can be efficiently parallelized on modern multi-core computers; the C part
also parallelizable on clusters.

The {\tt FIESTA} user interface is based on {\tt
Mathematica}. To run {\tt FIESTA}, the user has to load the {\tt
FIESTA\_2.0.0.m} into {\tt  Mathematica} 6 or 7. In order to evaluate a
Feynman integral one has to use the command
\begin{alltt}
{\tt SDEvaluate[{U,F,\(\ell\)},indices,order]},
\end{alltt}
where {\tt U} and {\tt F} are the functions from formula
(\ref{Alpha}), {$\ell$} is the number of loops,
{\tt indices} is the set of indices and {\tt order} is the required
order of the $\ep$-expansion.

To avoid manual construction of $U$ and $F$ one can use a build-in function {\tt UF}
and launch the evaluation as follows:
\begin{alltt}
{\tt SDEvaluate[UF[loop\_momenta,propagators,subst],indices,order]},
\end{alltt}
where {\tt subst} is a set of substitutions for external momenta, masses and
other values (please note that the code performs numerical integrations, therefore the
functions {\tt U} and {\tt F} should not depend on any external kinematic invariants).

Example:\\[1em]
\hspace*{1ex}{\tt SDEvaluate[UF[\{k\},\{-k$^2$,-(k+p$_1$)$^2$,-(k+p$_1$+p$_2$)$^2$,-(k+p$_1$+p$_2$+p$_4$)$^2$\},
\\
\hspace*{1ex}\{p$_1^2\rightarrow$0,p$_2^2\rightarrow$0,p$_4^2\rightarrow$0,
p$_1$ p$_2\rightarrow$-s/2,p$_2$ p$_4\rightarrow$-t/2,p$_1$ p$_4\rightarrow$-(s+t)/2,
\\
\hspace*{1ex}s$\rightarrow$-3,t$\rightarrow$-1\}],
\{1,1,1,1\},0]
}\\[1em]
performs an evaluation of the massless on-shell box diagram
where the Mandelstam variables are equal to  $s=-3$ and $t=-1$.

\section{Numerical results for four-loop massless propagators}

\ice{
\begin{figure}[!hbt]
\begin{center}
\includegraphics[width=.9\textwidth]{m41-63.eps}
\caption{Most complicated four-loop propagator diagrams}
\label{m41-63}
\end{center}
\end{figure}
}

\begin{figure}[!hbt]
\begin{center}
\SetScale{0.8}
\SetWidth{1.0}

\begin{picture}(75,90)(0,0)
\CArc(50,50)(20,0,360)
\Line(70,50)(80,50)
\Line(30,50)(20,50)
\Line(61,67)(61,33)
\Line(39,67)(39,34)
\Line(39,50)(61,50)
\put(33,8){$M_{61},\,\ep^1$}
\end{picture}
\begin{picture}(75,90)(0,0)
\CArc(50,50)(20,0,360)
\Line(70,50)(80,50)
\Line(30,50)(20,50)
\Line(62,34)(62,47)
\Line(38,34)(38,47)
\Line(38,47)(62,47)
\Line(62,47)(36,64.3333333333333)
\Line(38,47)(46.3205029433784,52.5470019622523)
\Line(64,64)(53,56.6666666666667)
\put(33,8){$M_{62},\,\ep^0$}
\end{picture}
\begin{picture}(75,90)(0,0)
\CArc(50,50)(20,0,360)
\Line(70,50)(80,50)
\Line(30,50)(20,50)
\Line(39,67)(39,33)
\Line(39,50)(68,57.25)
\Line(61,33)(61,53)
\Line(61,67)(61,58)
\put(33,8){$M_{63},\,\ep^0$}
\end{picture}
\begin{picture}(75,90)(0,0)
\CArc(50,50)(20,0,360)
\Line(70,50)(80,50)
\Line(30,50)(20,50)
\Line(50,70)(34,38)
\Line(40,50)(65,37.5)
\Line(50,70)(50,48)
\Line(50,30)(50,42)
\put(33,8){$M_{51},\,\ep^1$}
\end{picture}
\begin{picture}(75,90)(0,0)
\CArc(50,50)(20,0,360)
\Line(61,67)(52,53.5)
\Line(39,34)(48,47.5)
\Line(50,50)(39,66.5)
\Line(50,50)(61,33.5)
\Line(61,67)(39,67)
\Line(70,48)(80,48)
\Line(30,48)(20,48)
\put(33,8){$M_{41},\,\ep^1$}
\end{picture}

\begin{picture}(75,90)(0,0)
\CArc(50,50)(20,0,360)
\Line(61,66)(52,52.5)
\Line(39,34)(48,47.5)
\Line(50,50)(39,66.5)
\Line(50,50)(61,33.5)
\Line(61,66)(70,48)
\Line(70,48)(80,48)
\Line(30,48)(20,48)
\put(33,8){$M_{42},\,\ep^1$}
\end{picture}
\begin{picture}(75,90)(0,0)
\CArc(50,50)(20,0,360)
\Line(50,50)(61,66.5)
\Line(50,50)(39,33.5)
\Line(70,48)(80,48)
\Line(30,48)(20,48)
\Line(61,67)(61,33)
\Line(39,67)(39,34)
\put(33,8){$M_{44},\,\ep^0$}
\end{picture}
\begin{picture}(75,90)(0,0)
\CArc(50,50)(20,0,360)
\Line(61,66)(52,52.5)
\Line(39,34)(48,47.5)
\Line(50,50)(39,66.5)
\Line(50,50)(61,33.5)
\Line(61,67)(61,34)
\Line(70,48)(80,48)
\Line(30,48)(20,48)
\put(33,8){$M_{45},\,\ep^1$}
\end{picture}
\begin{picture}(75,90)(0,0)
\Line(30,50)(20,50)
\Line(70,50)(80,50)
\CArc(50,50)(20,0,360)
\Line(34,62)(66,62)
\Line(50,30)(34,62)
\Line(50,30)(66,62)
\put(33,8){$M_{34},\,\ep^3$}
\end{picture}
\begin{picture}(75,90)(0,0)
\Line(30,50)(20,50)
\Line(50,50)(80,50)
\CArc(50,50)(20,0,360)
\Line(50,50)(41,68)
\Line(50,50)(41,32)
\Line(41,32)(41,68)
\put(33,8){$M_{35},\,\ep^2$}
\end{picture}

\begin{picture}(75,90)(0,0)
\Line(20,50)(80,50)
\CArc(50,50)(20,0,360)
\Line(50,30)(50,70)
\put(33,8){$M_{36},\,\ep^1$}
\end{picture}\begin{picture}(75,90)(0,0)
\CArc(40,50)(10,0,360)
\CArc(66,50)(15,0,360)
\Line(55,61)(77,39)
\Line(77,61)(67.54,51.54)
\Line(64.46,48.46)(55,39)
\Line(30,50)(20,50)
\Line(82,50)(92,50)
\put(33,8){$M_{52},\,\ep^1$}
\end{picture}
\begin{picture}(75,90)(0,0)
\CArc(50,50)(20,0,360)
\Line(20,50)(80,50)
\Line(42,68)(60,50)
\Line(58,68)(51.6496,61.6496)
\Line(47.56,57.56)(40,50)
\put(33,8){$M_{43},\,\ep^1$}
\end{picture}
\begin{picture}(75,80)(-15,-10)
\SetScale{1.0}
\SetWidth{0.8}
\put(15,0){$N_0,\ep^2$}
\CArc(26,30)(15,0,360)
\Line(15,41)(37,19)
\Line(37,41)(28,32)
\Line(15,19)(23.4852813742386,27.4852813742386)
\Line(10,30)(5,30)
\Line(42,30)(47,30)
\end{picture}
\caption{ $M_{61}$ ---  $M_{43}$: the thirteen   complicated four-loop master integrals according to \cite{BaCh}.
The integrals are ordered  (if  read from left to right and then from top to bottom) according 
to their complexity. The two MI's $M_{52}$ and $M_{43}$ can be identically  expressed through  the  three-loop nonplanar MI $N_0$.  
}
\label{m41-63}
\end{center}
\end{figure}
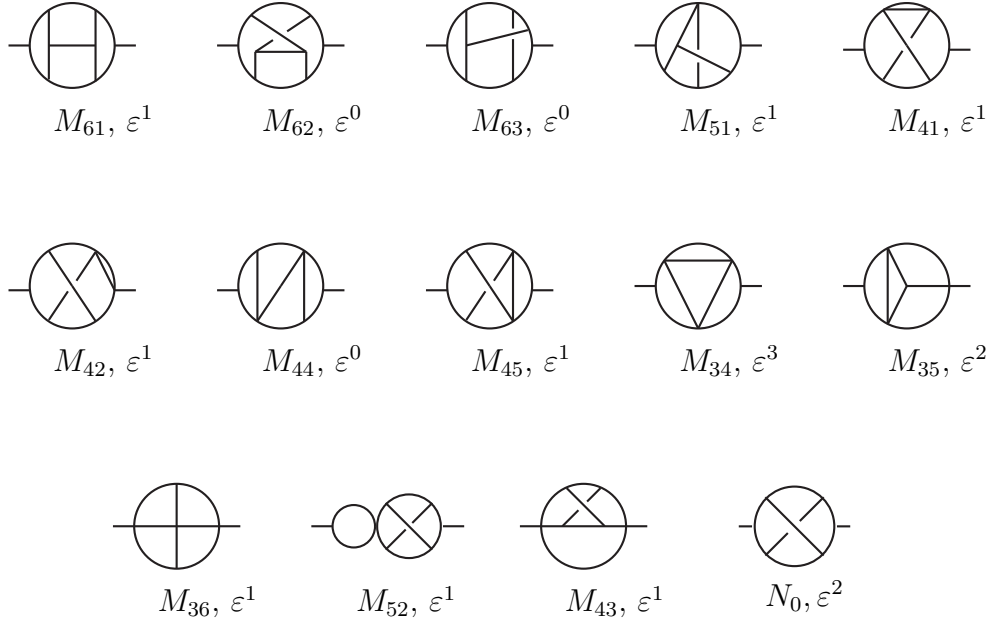

In \cite{BaikovCriterion} a  full set of the
four-loop massless propagator-like  MI's was identified.
There exist 28 independent MI's. Analytical results for these
integrals were obtained in
\cite{BaCh}.  By an analytical result is meant {\em  not}
an analytical expression for a master integral taken at a generic value
of the  space-time dimension $d$ (which is usually not possible
except for the simplest cases), but rather the analytic expressions  for {\em proper}
number of terms in its Laurent expansion in $d$ around the physical
value $d=4$.  

As it was shown in \cite{BaCh}, the full set of all 28 integrals can be
divided in three parts: 
\begin{enumerate}
\item 9 ``primitive'' integrals which are expressible in terms of
  $\Gamma$-functions;
\item 6 ``simple'' integrals which could be expressed in terms of
  $\Gamma$-functions and the so-called generalized two-loop diagram
  with insertions;
\item remaining 13 ``complicated'' integrals are quite difficult for
  both analytical and numerical evaluation.
\end{enumerate}

The complicated MI's are pictured in Fig.~\ref{m41-63}. The MI's are
labeled as $M_{ij}$ where the first index, $i$, stands for the number
of internal lines minus five while the second index, $j$, numerates
(starting from 1) different integrals with the same $i$.
$\ep^{\scriptsize m}$ after $M_{ij}$ stands for the maximal term in
$\ep$-expansion of $M_{ij}$ which one needs to know for evaluation of
the contribution of the integral to the final result for a four-loop
integral after reduction is done, see \cite{BaCh}. That is, $m$ stands for the
maximal power of a spurious pole $1/\ep^m$ which could appear in front
of $M_{ij}$ in the process of reduction to masters.

Primitive and simple integrals are known analytically. 
Two of the complicated integrals ($M_{43}$ and $M_{52}$) are related by
a simple factor with the three-loop MI $N_0$  \cite{BaCh} so it is enough to evaluate remaining
eleven complicated MI's   $M_{61}$ ---  $M_{36}$ as well as first three terms  of the $\ep$-expansion of $N_0$. 

We have calculated all them by means of {\tt FIESTA}.
We used the {\tt Cuba} {\tt Vegas}
integrator with different parameters used for the numerical
integration. Evaluations were performed on 
8-core (2x4) Intel Xeon E5472 3.0 GHz, 4GB/core RAM,
4.6TB disk/node computers in fully parallel mode, i.e., both
Mathematica and C parts were completely parallelized. The square of
the external momentum $q$ was chosen as -1: $q^2 = -1$. The
{\tt FIESTA} input for, say,  the integral $M_{44}$ reads:
\begin{verbatim}
Fm44= {
   -k1^2,  -k3^2, -k4^2, -(k1+q)^2, -(k2+q)^2, -(k4+q)^2,
   -(k1-k2)^2, -(k2-k3)^2, -(k3-k4)^2
};

SDEvaluate[UF[{k1,k2,k3,k4}, Fm44,{q^2-> -1}],
   {1,1,1,1,1,1,1,1,1},1];
\end{verbatim}

Our results alongside with the corresponding analytical expressions
(transformed to the numerical form) from \cite{BaCh} are presented in
Tables 1 and 2.

Within {\tt FIESTA} it is implied that Feynman integrals are with the
$-k^2-i0$ dependence of propagators and results are presented, in a
Laurent expansion in $\ep$, by pulling out the factor
$i\pi^{d/2}e^{-\gamma_E\ep}$ per loop, where $\gamma_E$ is the
Euler constant. Please, note that the overall normalization used by
{\tt FIESTA} is {\em different} from the one employed by the authors of
\cite{BaCh}.  We denote by $\ovl{M}_i$ a {\tt FIESTA} result for an $\ell$-loop  MI
$M_i$.
The connection between both values reads:
\[
\ovl{M}_i = \left[ e^{\gamma_E\ep}\,\frac{\Gamma(1 +\ep)\Gamma(1-\ep)^2}{\Gamma(2 -2\ep)}\right]^{\ell}\, M_i
{}.
\]

Numerically, for $\ell=4 $  the  conversion factor is: 
\bea \nonumber
\ovl{M}_i &=&   \left[ 1 + 8 \,\ep + 36.710132 \,\ep^2 + 122.46185717 \,\ep^3 + 
 329.99310668\,\ep^4 + 
\right.
\\
&+&
\left.
758.778374 \,\ep^5 + 1543.7276075 \,\ep^6 + 
 2848.0962405 \,\ep^7 +{\cal O}(\ep^8)\right]  M_i
\eea
and for $\ell=3$ 
\beq
\ovl{N}_0 = \left[ 1 + 6\,\ep + 21.53259889972766\,\ep^2 +{\cal O}(\ep^3)\right]  N_0.
\eeq

\begin{table}[p]
\begin{minipage}{\linewidth}
\begin{tabular}{|c|c|c|c|c|}
\hline
Int. & Degree & Exact & Cuba Vegas & Cuba Vegas\\
id:&of $\ep$:&Value:&500 000 result:&1 500 000 result:\\
\hline
\hline 
    &  $\ep^{-4}$ & 0.08333   & 0.08333 $\pm$ 0 & 0.08333 $\pm$ 0\\
    &  $\ep^{-3}$ & 0.91666  & 0.916668 $\pm$ 0.00003 & 0.916667 $\pm$ 0.000018\\
    &  $\ep^{-2}$ & 5.6425109  & 5.64252 $\pm$ 0.00038 &  5.64251 $\pm$ 0.00022\\
    &  $\ep^{-1}$ & 27.6412581  & 27.6413 $\pm$ 0.0013 & 27.6413 $\pm$ 0.00077\\
$\ovl{M}_{34}$ &  $\ep^{0}$  & 98.637928  & 98.638 $\pm$ 0.0058 & 98.638 $\pm$ 0.0034\\
    &  $\ep^{1}$  & 342.7349920 & 342.738 $\pm$ 0.021 & 342.736 $\pm$ 0.012\\
    &  $\ep^{2}$  & 857.8735165  & 857.88 $\pm$ 0.081 & 857.88 $\pm$ 0.048\\
    &  $\ep^{3}$  & 2659.825402  & 2659.86 $\pm$ 0.32 &  2659.84 $\pm$ 0.19\\
    &  $\ep^{4}$  & ---  & 4344.27 $\pm$ 1.3 & 4344.28 $\pm$ 0.75\\
\hline
\hline
    &  $\ep^{-2}$ & 0.601028  & 0.601030 $\pm$ 0.000024& 0.601028 $\pm$ 0.000012\\
    &  $\ep^{-1}$ & 7.4230554  & 7.4232 $\pm$ 0.0004& 7.4231 $\pm$ 0.00024\\
$\ovl{M}_{35}$ &  $\ep^{0}$ & 44.91255  & 44.9128 $\pm$ 0.0012& 44.9127 $\pm$ 0.00073\\
    &  $\ep^{1}$ & 217.0209011  & 217.026 $\pm$ 0.0062 & 217.023 $\pm$ 0.0037\\
    &  $\ep^{2}$ & 780.4321125  & 780.439 $\pm$ 0.022& 780.436 $\pm$ 0.013\\
    &  $\ep^{3}$ &  --- & 2678.18 $\pm$ 0.09 & 2678.13 $\pm$ 0.053\\
\hline
\hline
    &  $\ep^{-1}$ & 5.1846388  &5.18467 $\pm$ 0.000072 & 5.184645 $\pm$ 0.000042\\
$\ovl{M}_{36}$ &  $\ep^{0}$ & 38.8946741  & 38.8950 $\pm$ 0.00068& 38.8948 $\pm$ 0.00039\\
    &  $\ep^{1}$ & 240.0684359  &240.071 $\pm$ 0.0032 & 240.069 $\pm$ 0.0019\\
    &  $\ep^{2}$ &  --- &948.630 $\pm$ 0.016 & 948.623 $\pm$ 0.0091\\
\hline
\hline
    &  $\ep^{-1}$ & 20.7385551  & 20.7386 $\pm$ 0.0004& 20.73860 $\pm$ 0.00023\\
$\ovl{M}_{41}$ &  $\ep^{0}$ & 102.0326759  & 102.034 $\pm$ 0.0051& 102.033 $\pm$ 0.003\\
    &  $\ep^{1}$ & 761.5969858  & 761.61 $\pm$ 0.019& 761.60 $\pm$ 0.011\\
    &  $\ep^{2}$ &  ---  & 2326.21 $\pm$ 0.11& 2326.18 $\pm$ 0.062\\
\hline
\end{tabular}
\end{minipage}
\caption{\label{BaikovChetIntsVal} Numerical results for the MI's. In the third column
the numerical values of the known analytical results are shown. The last two columns 
contain the results of evaluation on these integrals by {\tt FIESTA}
using the {\tt Cuba} {\tt Vegas} integrator
with 500000 and 1500000 sampling points correspondingly. For all MI's we
calculate one extra $\ep$-term (not known analytically).}
\end{table}

\begin{table}[p]
\begin{minipage}{\linewidth}
\begin{tabular}{|c|c|c|c|c|}
\hline
Int. & Degree & Exact & Cuba Vegas & Cuba Vegas\\
id:&of $\ep$:&Value:&500 000 result:&1 500 000 result:\\
\hline
\hline 
    &  $\ep^{-1}$ & 20.7385551  & 20.7386 $\pm$ 0.00041& 20.73860 $\pm$ 0.00024\\
$\ovl{M}_{42}$ &  $\ep^{0}$ & 145.3808999  & 145.382 $\pm$ 0.0049& 145.381 $\pm$ 0.0029\\
    &  $\ep^{1}$ & 985.9082306  & 985.92 $\pm$ 0.023& 985.91 $\pm$ 0.014\\
    &  $\ep^{2}$ & ---  & 3930.68 $\pm$ 0.13& 3930.65 $\pm$ 0.076\\
\hline
\hline
$\ovl{M}_{44}$ &  $\ep^{0}$ & 55.5852539&55.5858 $\pm$ 0.00054 &55.58537 $\pm$ 0.00031  \\
    &  $\ep^{1}$ & --- & 175.325 $\pm$ 0.006 &175.325 $\pm$ 0.004  \\
\hline
\hline
    &  $\ep^{0}$ & 52.0178687  & 52.0184 $\pm$ 0.00052& 52.0181 $\pm$ 0.0003\\
$\ovl{M}_{45}$ &  $\ep^{1}$ & 175.496447  & 175.50 $\pm$ 0.0062& 175.50 $\pm$ 0.0036\\
    &  $\ep^{2}$ & ---  & 1475.29 $\pm$ 0.017& 1475.272 $\pm$ 0.0098\\
\hline
\hline
    &  $\ep^{-1}$ & -5.1846388  & -5.18466 $\pm$ 0.000081& -5.184651 $\pm$ 0.000048\\
$\ovl{M}_{51}$ &  $\ep^{0}$ & -32.096143  & -32.0966 $\pm$ 0.00097& -32.0962 $\pm$ 0.00057\\
    &  $\ep^{1}$ & -91.1614758  & -91.157 $\pm$ 0.009& -91.158 $\pm$ 0.0052\\
    &  $\ep^{2}$ & ---  & 119.08 $\pm$ 0.074& 119.06 $\pm$ 0.043\\
\hline
\hline
    &  $\ep^{0}$ & 20.7385551  & 20.7387 $\pm$ 0.00045& 20.73857 $\pm$ 0.00026\\
$\ovl{N}_0$
&  $\ep^{1}$ & 190.600238  & 190.60 $\pm$ 0.004& 190.60 $\pm$ 0.0023\\
    &  $\ep^{2}$ & 1049.194196  & 1049.20 $\pm$ 0.025& 1049.20 $\pm$ 0.014\\
    &  $\ep^{3}$ & ---  & 4423.86 $\pm$ 0.12& 4423.84 $\pm$ 0.072\\
\hline
\hline
    &  $\ep^{-1}$ & -10.3692776  & -10.36941 $\pm$ 0.00011& -10.36931 $\pm$ 0.00006\\
$\ovl{M}_{61}$ &  $\ep^{0}$ & -70.99081719  & -70.989 $\pm$ 0.002& -70.990 $\pm$ 0.0011\\
    &  $\ep^{1}$ & -21.663005  & -21.633 $\pm$ 0.023& -21.650 $\pm$ 0.013\\
    &  $\ep^{2}$ & ---  & 2832.86 $\pm$ 0.17 & 2832.69 $\pm$ 0.096(\footnote{
Calculated with the {\tt Fortran} {\tt Vegas} using 1 550 000 samples.})\\
\hline
\hline
    &  $\ep^{-1}$ & -10.3692776  & -10.36940 $\pm$ 0.00011& -10.36933 $\pm$ 0.00006\\
$\ovl{M}_{62}$ &  $\ep^{0}$ & -58.6210462  & -58.6174 $\pm$ 0.0022& -58.6187 $\pm$ 0.0013\\
    &  $\ep^{1}$ & ---  & 244.69 $\pm$ 0.025& 244.681 $\pm$ 0.015\\
\hline
\hline
    &  $\ep^{-1}$ & -5.1846388  &-5.18470 $\pm$ 0.000078 & -5.18467 $\pm$ 0.000042\\
$\ovl{M}_{63}$ &  $\ep^{0}$ & 14.397395  &14.40 $\pm$ 0.0014 & 14.3989 $\pm$ 0.00081\\
    &  $\ep^{1}$ & ---  & 740.00 $\pm$ 0.017& 739.979 $\pm$ 0.0099\\
\hline
\end{tabular}
\end{minipage}
\caption{\label{BaikovChetIntsValCont}
Continuation of the table \ref{BaikovChetIntsVal}.}
\end{table}

\begin{table}[p]
\begin{minipage}{\linewidth}
\begin{tabular}{|c|c|r@{.}l|r@{.}l|}
\hline
Int. & Degree &      \multicolumn{2}{|c|}{Cuba Vegas} & \multicolumn{2}{|c|}{Cuba Vegas}\\
id:&of $\ep$:&\multicolumn{2}{|c|}{500 000 time:}&\multicolumn{2}{|c|}{1 500 000 time:}\\
\hline
\hline 
    &  $\ep^{-4}$ & 60&77s& 59&37s\\
    &  $\ep^{-3}$ & 63&56s& 65&95s\\
    &  $\ep^{-2}$ & 82&89s& 127&82s\\
    &  $\ep^{-1}$ & 211&84s& 521&50s\\
$\ovl{M}_{34}$ &  $\ep^{0}$  & 401&53s& 1064&87s\\
    &  $\ep^{1}$  & 586&88s& 1608&08s\\
    &  $\ep^{2}$  & 988&27s& 2739&44s\\
    &  $\ep^{3}$  & 1870&97s& 5225&72s\\
    &  $\ep^{4}$  & 3422&80s& 9572&06s\\
    \cline{2-6}
    &  Total time: & 9847& 39s& 23163&80s\\
\hline
\hline
    &  $\ep^{-2}$ & 64&29s &76 &14s\\
    &  $\ep^{-1}$ & 178&16s &426 &75s\\
$\ovl{M}_{35}$ &  $\ep^{0}$  & 325&72s &855 &19s\\
    &  $\ep^{1}$  & 436&57s &1169 &30s\\
    &  $\ep^{2}$  & 678&91s &1828 &51s\\
    &  $\ep^{3}$  & 1275&64s &3532 &42s\\
    \cline{2-6}
    &  Total time: &3764&31s&8694&65s\\
\hline
\hline
    &  $\ep^{-1}$ & 152&96s &375 &47s\\
$\ovl{M}_{36}$ &  $\ep^{0}$  &269 &92s &704 &23s\\
    &  $\ep^{1}$  &354 &42s &959 &94s\\
    &  $\ep^{2}$  &526 &79s &1442 &97s\\
    \cline{2-6}
    &  Total time: & 1590&61s &3769 &25s\\
\hline
\hline
    &  $\ep^{-1}$ &185 &32s & 274&82s\\
$\ovl{M}_{41}$ &  $\ep^{0}$  &691 &48s &1764 &69s\\
    &  $\ep^{1}$  &928 &23s &2431 &03s\\
    &  $\ep^{2}$  &1260 &22s &3379 &72s\\
    \cline{2-6}
    &  Total time: &3776 &42s &8562 &06s\\
\hline
\end{tabular}
\end{minipage}
\caption{\label{BaikovChetIntsTime} Timing for calculations of the MI's. The last two columns contain time (in seconds) of numerical
  integration by the {\tt Cuba} {\tt Vegas} integrator with 500000 and 1500000
  sampling points. Also a total time for evaluation of each integral
  is given, including the Mathematica part.}
\end{table}

\begin{table}[p]
\begin{minipage}{\linewidth}
\begin{tabular}{|c|c|r@{.}l|r@{.}l|}
\hline
Int. & Degree &      \multicolumn{2}{|c|}{Cuba Vegas} & \multicolumn{2}{|c|}{Cuba Vegas}\\
id:&of $\ep$:&\multicolumn{2}{|c|}{500 000 time:}&\multicolumn{2}{|c|}{1 500 000 time:}\\
\hline
\hline
    &  $\ep^{-1}$ &176 &02 & 246&99s\\
$\ovl{M}_{42}$ &  $\ep^{0}$  &686 &57 & 1762&30s\\
    &  $\ep^{1}$  &917 &95 & 2435&75s\\
    &  $\ep^{2}$  &1233 &20 & 3289&26s\\
    \cline{2-6}
    &  Total time: &3753 &92s &8485 &04s\\
\hline
\hline
$\ovl{M}_{44}$ &  $\ep^{0}$  & 798&20s &2097 &39s\\
    &  $\ep^{1}$  & 1016&19s &2713 &11s\\
    \cline{2-6}
    &  Total time: & 2174&60s &5185 &72s\\
\hline
\hline
    &  $\ep^{0}$  & 750&16s &1906 &93s\\
$\ovl{M}_{45}$ &  $\ep^{1}$  & 975&67s &2533 &32s\\
    &  $\ep^{2}$  & 1246& 26s& 3256&41s\\
    \cline{2-6}
    &  Total time: & 3713&05s &8416 &63s\\
\hline

\hline
    &  $\ep^{-1}$ & 516&89s &698 &85\\
$\ovl{M}_{51}$ &  $\ep^{0}$  & 1676&80s &4206 &21\\
    &  $\ep^{1}$  & 2881&73s &7672 &50\\
    &  $\ep^{2}$  & 3597&15s & 9615&16s\\
    \cline{2-6}
    &  Total time: & 10736&08s & 24277&30s\\
\hline
\hline
    &  $\ep^{0}$  & 42&13s &104 &29s\\
$\ovl{N}_0$ &  $\ep^{1}$  & 75& 01s&201 &78s\\
    &  $\ep^{2}$  & 90&57s &246 &34s\\
    &  $\ep^{3}$  & 129& 84s&341 &99s\\
    \cline{2-6}
    &  Total time: & 411&12s &967 &20s\\
\hline
\hline
    &  $\ep^{-1}$ & 2262&86s &3495 &28s\\
$\ovl{M}_{61}$ &  $\ep^{0}$  & 15242&10s &39673 &48s\\
    &  $\ep^{1}$  & 61481&36s &162453 &52s\\
    &  $\ep^{2}$  & 202018&31s&1794640&00s(\footnote{
Integration by the {\tt Fortran} {\tt Vegas} using 1 550 000 samples.})\\
    \cline{2-6}
    &  Total time: &768727 &00s &\multicolumn{2}{|c|}{---}\\
\hline
\hline
    &  $\ep^{-1}$ & 3003&05s &4131 &07s\\
$\ovl{M}_{62}$ &  $\ep^{0}$  & 16073&09s &39690 &66s\\
    &  $\ep^{1}$  & 63720&52s &163026 &12s\\
    \cline{2-6}
    &  Total time: & 156510&00s &280778 &00s\\
\hline
\hline
    &  $\ep^{-1}$ & 273900& 44s& 3316&17s\\
$\ovl{M}_{63}$ &  $\ep^{0}$  & 14870&92s &36434 &93s\\
    &  $\ep^{1}$  & 59206&88s &151788 &20s\\
    \cline{2-6}
    &  Total time: & 147870& 00s& 262670&00s\\
\hline
\end{tabular}
\end{minipage}
\caption{\label{BaikovChetIntsTimeCont}Continuation of the table \ref{BaikovChetIntsTime}.}
\end{table}

A comparison with the analytical results 
shows that the integration with \mbox{500 000} sampling points leads
to the numerical result with 3-4 reliable digits in a quite reasonable
time (see the tables \ref{BaikovChetIntsTime} and
\ref{BaikovChetIntsTimeCont}) while the integration with 
\mbox{1 500  000} sampling points reproduces the analytical results with 4-5
digits. We have also evaluated one extra term in the $\ep$-expansion
of each MI which is currently unavailable analytically but is
necessary for future five-loop calculations.  For some technical 
reasons\footnote{
The evaluation was performed before we've implemented the {\tt Cuba}
library. The integrator spends 1794640 seconds which is more than 20 days so we
wouldn't like to load 8-core machine for such a period by the job
which was already done},
for the highest $\ep$ term of the
integral $\ovl{M}_{61}$, we restricted ourselves with the value produced
by the {\tt Fortran} {\tt Vegas} integrator which is not supported anymore.

Surprisingly this planar integral ($\ovl{M}_{61}$) appears to be the  most complicated
one for numerical integration, see the table
\ref{BaikovChetIntsTimeCont}. 
Non-planar integrals $\ovl{M}_{62}$ and $\ovl{M}_{63}$ are
also complicated for {\tt FIESTA} but much less than ${M}_{61}$. Other
integrals (including non-planars) are incomparably easier for
numerical evaluation by {\tt FIESTA}.

The first thirteen MI's from the Fig.~\ref{m41-63} are very difficult
for analytical evaluation, and only three of them had been checked in an
independent way, see \cite{BaCh}. If even one of the remaining ten
MI's was evaluated incorrectly, it would change all physical results
obtained with the use of these MI's.

Analytical results were obtained in \cite{BaCh} using so-called
``glue-and-cut'' symmetry together with the procedure of reduction to
MI's. The reduction procedure is extremely complicated, it
requires careful computer algebra programming and very large-scale
computer evaluations. In the present paper we have performed the
independent check of these MI's using completely different approach,
namely, sector decomposition, providing a quite strong evidence for
the correctness of the algorithms and their implementation in \cite{BaCh}.

\section{Conclusion}

Usually, analytical evaluation of multiloop MI is a kind of
art. It requires a lot of efforts (and sometimes CPU
time). In many situations, independent checkup is hardly any possible
in reasonable time. That is why the simple in use tools for numerical
evaluation like {\tt FIESTA} are  important.

Some of the integrals presented in this paper are really complicated,
and the original {\tt FIESTA 1} was not able to evaluate them at all.
This was one of our motivations, in particular, to improve {\tt FIESTA} so
that it would cope with these  (and, hopefully, many others) integrals. There had been 
both technical and theoretical complications which had to be  solved
\cite{FIESTA2} for this aim.

The successful check of the results of \cite{BaCh} demonstrates that
the current version of {\tt FIESTA} is a powerful tool for evaluating
integrals numerically and for cross-checking analytical results.

\vspace{0.2 cm}

{\em Acknowledgments.}
This work was supported in part by DFG through SBF/TR 9 and the
Russian Foundation for Basic Research through grant 08-02-01451.
We would like to thank K.~Chetyrkin and P.~Baikov for motivation,
fruitful discussions and attentive reading of the manuscript.

\end{document}